\documentclass[fleqn,usenatbib]{mnras}

\usepackage{newtxtext,newtxmath}
\usepackage{mathtools, cuted}

\usepackage[T1]{fontenc}

\DeclareRobustCommand{\VAN}[3]{#2}
\let\VANthebibliography\thebibliography
\def\thebibliography{\DeclareRobustCommand{\VAN}[3]{##3}\VANthebibliography}

\usepackage{graphicx}	
\usepackage{amsmath}	
\usepackage{amssymb}	

\title[explosion with point gravity]{Self Similar Adiabatic Strong Explosion in a Medium Gravitationally Free Falling to a Point Mass}

\author[Almog Yalinewich]{
Almog Yalinewich$^{1}$\thanks{E-mail: almog.yalin@gmail.com}
\\
$^{1}$Canadian Institute for Theoretical Astrophysics, 60 St. George St., Toronto, ON M5S 3H8, Canada
}

\date{Accepted XXX. Received YYY; in original form ZZZ}

\pubyear{2020}

\begin{document}
\label{firstpage}
\pagerange{\pageref{firstpage}--\pageref{lastpage}}
\maketitle

\begin{abstract}
We develop a generalisation to the classical Sedov Taylor explosion where the medium free falls to a point mass at the centre of the explosion. To verify our analytic results, we compare them to a suite of numerical simulations. We find that there exists a critical energy below which, instead of propagating outward the shock stalls and collapses under gravity. Furthermore, we find that the value of the critical energy threshold decreases when the adiabatic index increases and material is more evenly distributed within the shocked region. We apply this model to the problem of a shock bounce in core collapse supernova, in which the proto neutron star serves as the point mass. The relation between the threshold energy and the distribution of mass in the shock might help explain how turbulence prevents shock stalling and recession in a core collapse supernova explosion.
\end{abstract}




\section{Introduction}

Massive stars end their lives with a spectacular explosion called a supernova \citep[e.g.][and references therein]{VanDyk2017TheProgenitors}. Despite decades of research, this mechanism is not well understood \citep[e.g.][and references therein]{Janka2007TheorySupernovae, Muller2017TheMechanism}. At the end of its life, the star loses thermal pressure support and collapses  under its own gravity \citep{Burrows2012PerspectivesTheory}. This collapse triggers the release of a large amount of energy at the centre of the star, and gives rise to a shock wave that emerges from the centre of the star and, initially at least, moves outward. If the shock wave is strong enough to overcome gravity and make it all the way to the surface of the star, then there will be a supernova. If gravity overwhelms the shock, the latter stalls and recedes, and there will be failed supernova or a direct collapse to a black hole. The quality of an explosion that determines whether it endures or stalls and recedes is often called the explodability.

One of the difficulties in the analysis of the explosion mechanism in core collapse supernovae is its irreducible complexity. Many other astrophysical phenomena can, at least conceptually, be broken up into a series of simpler processes such that each can be analysed in isolation. In contrast, a realistic and successful supernova explosion requires the concurrent action of multiple physical processes such as gravity, nuclear reactions, radiation, neutrino transport and turbulence \citep{Couch2014TheExplosions, Mabanta2018HowExplosions}. One could argue that turbulence is the most difficult process to take into account, since it breaks spherical symmetry, and necessitates a three dimensional model.

The main tools used to study supernova explosions are numerical simulations. Three dimensional simulations \citep{Fryer2002ModelingDimensions, Takiwaki2011Three-dimensionalTransport, Ott2013General-relativisticSupernovae, Melson2015Neutrino-drivenConvection, Burrows2019Three-dimensionalStars} are considered the most reliable, but are very resource intensive, and take a long time to run \citep{Mabanta2019Convection-aidedValidation}. Two dimensional simulations \citep{Livne2003Two-dimensionalContext, Buras2005Two-dimensionalStar, Ott20082DCores, Brandt2010ResultsTransport, Dolence2015Two-DimensionalTransport, Pan2016TWO-DIMENSIONALTRANSPORT} are less resource intensive on the one hand, but on the other hand are physically different from three dimensional simulations \citep{Takiwaki2014ASupernovae, Couch2014TheExplosions}. The most salient difference is the turbulence, which in two dimension has an inverse cascade, a feature absent from three dimensional turbulence \citep{Radice2018TurbulenceSupernovae}. For this reason it is not clear how to extrapolate the two dimensional results to the three dimensional case. Some authors even employ one dimensional simulations where the multi dimensional effect are included as extra terms in the flow equation \citep{Ugliano2012Progenitor-explosionProgenitors, Murphy2011ASupernovae, Mabanta2019Convection-aidedValidation, Couch2020SimulatingDimension}. These terms do not come from first principles, but are empirically calibrated from simulations in higher dimensions. 

These simulations are essential for understanding the explosion mechanism, but they all suffer from similar issues. First, it has been shown that small changes in the parameters can cause a big change in the outcome, e.g. explodability \citep{OConnor2017Core-CollapseEOS} and relic\footnote{Some authors refer to the compact object left behind a supernova as a remnant, but since this can be confused with a supernova remnant, here we refer to the former as relic.} mass \citep{Woosley2007NucleosynthesisMetallicity, Couch2020SimulatingDimension}. Second, different computational models yield different results for the same initial conditions \citep{OConnor2018GlobalSymmetry, Cabezon2018Core-collapseProject}. 

In this paper we propose a new approach to study supernova shock waves using self similar solutions \citep{Landau1987FluidMechanics, Barenblatt2014ScalingAsymptotics}. In order to use this method, one must neglect many of the physical effects present at the cores of exploding stars. Despite this great reduction in complexity, self similar solutions can still reveal profound insights about the problem at hand, which are still relevant when the neglected effects are introduced. For this reason these solution are not meant to replace simulations, but to complement them. The model that we develop here involves the instantaneous release of energy at the core of the star at $t=0$ and point gravity from the proto - neutron star. Moreover, as will be discussed below, the initial gas density has to scale with distance to the centre as $r^{-2}$ for the self similar model to work. This is actually not too different from what is observed in simulations \citep[e.g. figure 3 in ][]{Mezzacappa2001SimulationMechanism}. We also assume that at time $t=0$ the star loses pressure support and begins to collapse.

We note that our approach is not the only analytic model developed to describe supernova shock waves, but it is different from previous endeavours. \cite{Murphy2015AnExplosions, Mabanta2018HowExplosions} developed an analytic model based on a single zone approximation for the shocked region. \cite{Coughlin2018WeakSupernovae} considered the propagation of weak shock waves in a hydrostatic stellar atmosphere. The motivation for that work was a failed supernova rather than a successful one, i.e. when not enough energy is released to unbind the star. In contrast, the model described here assumes a strong shock, and that the stellar material is free falling rather than in hydrostatic equilibrium. \cite{Chevalier1989NeutronSupernova} and \cite{Kazhdan1994Self-similarMass} employed a self similar solution with a finite Mach number to model the accretion shock into the proto neutron star. In our picture, if such shock does form, it will be interior to the supernova shock wave. 

The plan of the paper is as follows. In section \ref{sec:collapse} we describe the evolution of the upstream material from the moment it begins to free fall to the point it is swept by the shock. In section \ref{sec:self_similar_solution} we present the self similar solution and its verification using a time dependent, finite volume hydrodynamic simulation. Finally, in section \ref{sec:discussion} we discuss the results and their implications for core collapse supernovae.

\section{The Collapse} \label{sec:collapse}

Let us assume some pre - existing, static density profile $\rho_a \left(r_0\right)$, where $r_0$ is the initial distance at time $t=0$ between some fluid element and the point mass. Initially, it is supported against the gravity of the central object $M$ by some thermal pressure gradient. At time $t = 0$ all the material cools immediately (or at least on a time much shorter than the Keplerian time). As a result, all fluid elements fall freely toward the central mass $M$. The trajectory of a single fluid element originally at $r_0$ is given implicitly by \citep{Lemos1989AGravity}
\begin{equation}
    \sqrt{\frac{2 \mathcal{G} M}{r_0^3}} t = \sqrt{\frac{r}{r_0} \left(1-\frac{r}{r_0}\right)} + \cos^{-1} \left(\sqrt{\frac{r}{r_0}}\right)
\end{equation}
where $r$ is the radius of some fluid element at time $t>0$ and $\mathcal{G}$ is the universal constant of gravity. Let us consider the material originally contained between radii $r_0$ and $r_0 + \Delta r_0$. At a later time the same material will occupy the region between $r$ and $r+\Delta r$, where the ratio between the thicknesses of the shells is given by
\begin{equation}
    \frac{\Delta r}{\Delta r_0} = \sigma \left(x\right) = \frac{3-x}{2} + \frac{3 \sqrt{1 - x} \cos^{-1}{\left(\sqrt{x} \right)}}{2 \sqrt{x}}
\end{equation}
where $x = r/r_0 < 1$. We note two extreme limits. In the first limit $\lim_{x\rightarrow 1} \sigma \left(x\right) = 1$ represents the case of a shell which has hardly moved. The other extreme limit represents the case where a shell is at a radius much smaller than its original radius $x\ll1$, in which case $\sigma \left(x\right) \approx \frac{3 \pi}{4\sqrt{x}}$. The mass enclosed between $r_0$ and $r_0 + \Delta r_0$ is $4 \pi r_0^2 \Delta r_0 \rho_a \left(r_0\right)$. Since mass is conserved, the density at a later time is given by

\begin{equation}
\rho = \rho_a \left(r_0\right) \frac{r_0^2 \Delta r_0}{r^2 \Delta r} = \frac{\rho_a \left(r/x\right)}{x^2 \sigma \left(x\right)}
\end{equation}
If the original density profile is some power law $\rho_a = k r^{-\omega}$ then at a later time the density profile is given by
\begin{equation}
    \rho = \frac{k r^{-\omega}}{x^{2-\omega} \sigma \left(x\right)} \, .
\end{equation}
The velocity of the shell is given by the conservation of energy
\begin{equation}
    v = \sqrt{\frac{2 \mathcal{G} M}{r} - \frac{2 \mathcal{G} M}{r_0}} = \sqrt{\frac{2 \mathcal{G} M}{r}} \sqrt{\frac{1}{x} - 1} \, .
\end{equation}

\section{Self Similar Solution} \label{sec:self_similar_solution}

In the absence of gravity, it is possible to obtain self similar solutions for explosions in any power law density profile $\rho_a = k r^{-\omega}$ (at least for $\omega<3$ so that the mass contained in the shocked region does not diverge) \citep{Rogers1957AnalyticDensity.}. In such cases, conservation of energy dictates that the shock radius scales with time as $R \propto t^{\frac{2}{5 -\omega}}$, or equivalently, that the shock speed scale with radius as $\dot{R} \propto R^{-\frac{3-\omega}{2}}$. However, when point gravity is introduce, we lose this degree of freedom. To preserve self similarity, the shock velocity has to be proportional to the keplerian velocity, so $\dot{R} \propto \sqrt{\mathcal{G} M/R}$ and hence $\omega = 2$. This means that a self similar solution can only be obtained for the case where the density profile scales as $\rho \propto r^{-2}$. The shock radius scales with time as $R \propto t^{2/3}$. This means that there's a constant ratio between the radius of the shock and the original radius of the shell it is sweeping $\chi = R/r_0$. Using this parameter we can write an explicit expression for the shock trajectory
\begin{equation}
    R = \varpi \left(\chi \right) \left(\mathcal{G} M t^2\right)^{1/3}
\end{equation}
where
\begin{equation}
    \varpi \left(\chi\right) = \sqrt[3]{2} \chi \left(\sqrt{\chi} \sqrt{1 - \chi} + \cos^{-1} {\left(\sqrt{\chi} \right)}\right)^{-\frac{2}{3}} \, .
\end{equation}
The relation between the shock velocity and the Keplerian velocity is
\begin{equation}
    \dot{R} = \frac{2}{3} \varpi^{3/2} \sqrt{\frac{\mathcal{G} M}{R}} \, .
\end{equation}
The density profile $\rho_a = k r_0^{-2}$ also defines a potential energy scale $\mathcal{G} M k$. The properties of the explosion will be determined by the ratio between the net energy of the explosion $E$ and the potential energy scale. We call this ratio the virial parameter $\epsilon = E/ 4 \pi \mathcal{G} M k$. We note that as the shock propagates the original potential energy of the swept up material does increase, but only logarithmically, so the ratio between explosion and potential energy can be assumed constant over many decades in radius.

In this section we will use these scaling relations to obtain the self similar profile of the hydrodynamic variables, and eventually relate the adiabatic index $\gamma$ and radius ratio $\chi$ to the virial parameter $\epsilon$.

\subsection{Differential Equations}

The hydrodynamic equations describing the interior of the shock \citep{Sakashita1974SimilarityFlow} are given by the conservation of mass

\begin{equation}
    \frac{\partial \rho}{\partial t} + \frac{1}{r^2} \frac{\partial}{\partial r} \left(r^2 \rho v\right) = 0
\end{equation}

the conservation of momentum 
\begin{equation}
    \frac{\partial v}{\partial t} + v \frac{\partial v}{\partial r} + \frac{1}{\rho} \frac{\partial p}{\partial r} = - \frac{\mathcal{G} M}{r^2}
\end{equation}

and the conservation of entropy
\begin{equation}
    \frac{\partial s}{\partial t} + v \frac{\partial s}{\partial r} = 0
\end{equation}
where $s = \ln p - \gamma \ln \rho$ is the entropy per unit mass. Using self similarity we can transform these partial differential equations to ordinary differential equations. We introduce the following dimensionless hydrodynamic variables
\begin{equation}
    \rho = k R^{-2} G\left(\xi\right)
\end{equation}
\begin{equation}
    v = \dot{R} V\left(\xi\right)
\end{equation}
\begin{equation}
    c = \dot{R} C \left(\xi\right)
\end{equation}
where $c = \sqrt{\gamma p/\rho}$ is the speed of sound, $R$ is the shock radius and $\xi = r/R$ is the dimensionless radius. Using these new variables, the hydrodynamic equations become \newpage
\begin{strip}
\begin{equation}
    \frac{G'\left(\xi\right)}{G\left(\xi\right)} = \frac{\left(8 \gamma \varpi^{3} \xi \left(\xi - V\right)^{3} + \gamma \left(\xi - V\right) \left(2 \varpi^{3} \xi^{2} V - 81\right) + \varpi^{3} \xi^{2} \left(12 - 8 \gamma\right) C^{2}\right)}{4 \varpi^{3} \xi^{2} \left(- \gamma \left(\xi - V\right)^{3} + \left(\xi - V\right) C^{2}{\left(\xi \right)} + \left(\gamma \xi - \gamma V - \xi + V\right) C^{2}\right)}
\end{equation}
\begin{equation}
    V'\left(\xi\right) = \frac{- 8 C^{2} V \gamma \varpi^{3} \xi + 12 C^{2} \varpi^{3} \xi^{2} - 2 V^{2} \gamma \varpi^{3} \xi^{2} + 2 V \gamma \varpi^{3} \xi^{3} + 81 V \gamma - 81 \gamma \xi}{4 \gamma \varpi^{3} \xi^{2} \left(C^{2} - V^{2} + 2 V \xi - \xi^{2}\right)}
\end{equation}
\begin{equation}
    C'\left(\xi\right) = \frac{C \left(8 \gamma \varpi^{3} \xi \left(V - \xi\right)^{2} \left(V \gamma - V - \gamma \xi + \xi\right) + \gamma \left(2 V \varpi^{3} \xi^{2} - 81\right) \left(V \gamma - V - \gamma \xi + \xi\right) - 4 \varpi^{3} \xi^{2} \left(C^{2} - \gamma \left(V - \xi\right)^{2}\right) \left(2 \gamma - 3\right)\right)}{8 \varpi^{3} \xi^{2} \left(C^{2} \left(V - \xi\right) + C^{2} \left(V \gamma - V - \gamma \xi + \xi\right) - \gamma \left(V - \xi\right)^{3}\right)}
\end{equation}
\end{strip}
The energy contained inside the shocked region is given by
\begin{equation}
    E = \int_0^{R} 4 \pi r^2 dr \rho \left(\frac{1}{2} v^2 + \frac{c^2}{\gamma \left(\gamma-1\right)} - \frac{\mathcal{G} M}{r}\right) = \label{eq:energy_equation}
\end{equation}
\begin{equation*}
    = \frac{16 \pi}{9} k \mathcal{G} M \varpi^3 \int_0^1 \xi^2 d \xi G \left(\frac{V^2}{2} + \frac{C^2}{\gamma\left(\gamma-1\right)} - \frac{9}{4 \varpi^3 \xi}\right)
\end{equation*}

\subsection{Boundary Conditions}

The relations between the conditions just ahead of the shock (upstream) and just behind the shock (downstream) are given by the strong shock Rankine Hugoniot conditions \citep{Zeldovich1967PhysicsPhenomena}
\begin{equation}
    \frac{\rho_d}{\rho_u} = \frac{\gamma+1}{\gamma-1}
\end{equation}
\begin{equation}
    \frac{v_d}{v_u} = \frac{\gamma-1}{\gamma+1}
\end{equation}
\begin{equation}
    p_d = \rho_u v_u \left(v_u-v_d\right)
\end{equation}
where $v_u$ is the upstream material velocity in the shock front frame, and $v_d$ is the downstream material velocity. The upstream values depend on $\chi$. The upstream density is given by
\begin{equation}
\rho_u = k R^{-2}/\sigma\left(\chi\right) \, .
\end{equation}
The downstream density is
\begin{equation}
    \rho_d = \frac{\gamma+1}{\gamma-1} \frac{k R^{-2}}{\sigma} \, .
\end{equation}
The shock is moving at $\dot{R}$ and the upstream fluid moves toward the shock at $\sqrt{\frac{2 \mathcal{G} M}{R} \left(\frac{1}{\chi} - 1\right)}$. Hence, the upstream velocity is 
\begin{equation}
    v_u = \dot{R} + \sqrt{\frac{2 \mathcal{G} M}{R} \left(\frac{1}{\chi}-1\right)} = \dot{R} \left(1-\frac{3 \sqrt{2}}{\varpi^{3/2}}\sqrt{\frac{1}{\chi}-1}\right)
\end{equation}
The downstream velocity is therefore
\begin{equation}
    v_d = \frac{\gamma-1}{\gamma+1} \dot{R} \left(1+\frac{3 \sqrt{2}}{\varpi^{3/2}}\sqrt{\frac{1}{\chi}-1}\right) \, .
\end{equation}
The downstream pressure is
\begin{equation}
    p_d = \frac{2}{\gamma+1} \frac{k R^2 \dot{R}^2}{\sigma} \left(1+\frac{3\sqrt{2}}{\varpi^{3/2}} \sqrt{\frac{1}{\chi}-1}\right)^2
\end{equation}
The downstream speed of sound is
\begin{equation}
    c_d = \sqrt{\gamma \frac{p_d}{\rho_d}} = \frac{\sqrt{2 \gamma \left(\gamma-1\right)}}{\gamma+1} \left(1+\frac{3 \sqrt{2}}{\varpi^{3/2}} \sqrt{\frac{1}{\chi} - 1}\right) \dot{R} \, .
\end{equation}
The only quantity that remains is the downstream velocity in the lab frame (where the point mass is stationary)
\begin{equation}
    v'_d = \dot{R} - v_d = \left(\frac{2}{\gamma+1} - \frac{\gamma-1}{\gamma+1} \frac{3 \sqrt{2}}{\varpi^{3/2}} \sqrt{\frac{1}{\chi}-1}\right) \dot{R} \, .
\end{equation}
The boundary conditions on the self similar variables are
\begin{equation}
    G\left(1\right) = \frac{\gamma+1}{\gamma-1} \frac{1}{\sigma}
\end{equation}
\begin{equation}
    C \left(1\right) = \frac{\sqrt{2 \gamma \left(\gamma-1\right)}}{\gamma+1} \left(1+\frac{3 \sqrt{2}}{\varpi^{3/2}} \sqrt{\frac{1}{\chi} - 1}\right)
\end{equation}
and
\begin{equation}
    V \left(1\right) = \frac{2}{\gamma+1} - \frac{\gamma-1}{\gamma+1} \frac{3 \sqrt{2}}{\varpi^{3/2}} \sqrt{\frac{1}{\chi}-1} \, .
\end{equation}

\subsection{Numerical Integration}

\begin{figure}
\includegraphics[width=\linewidth]{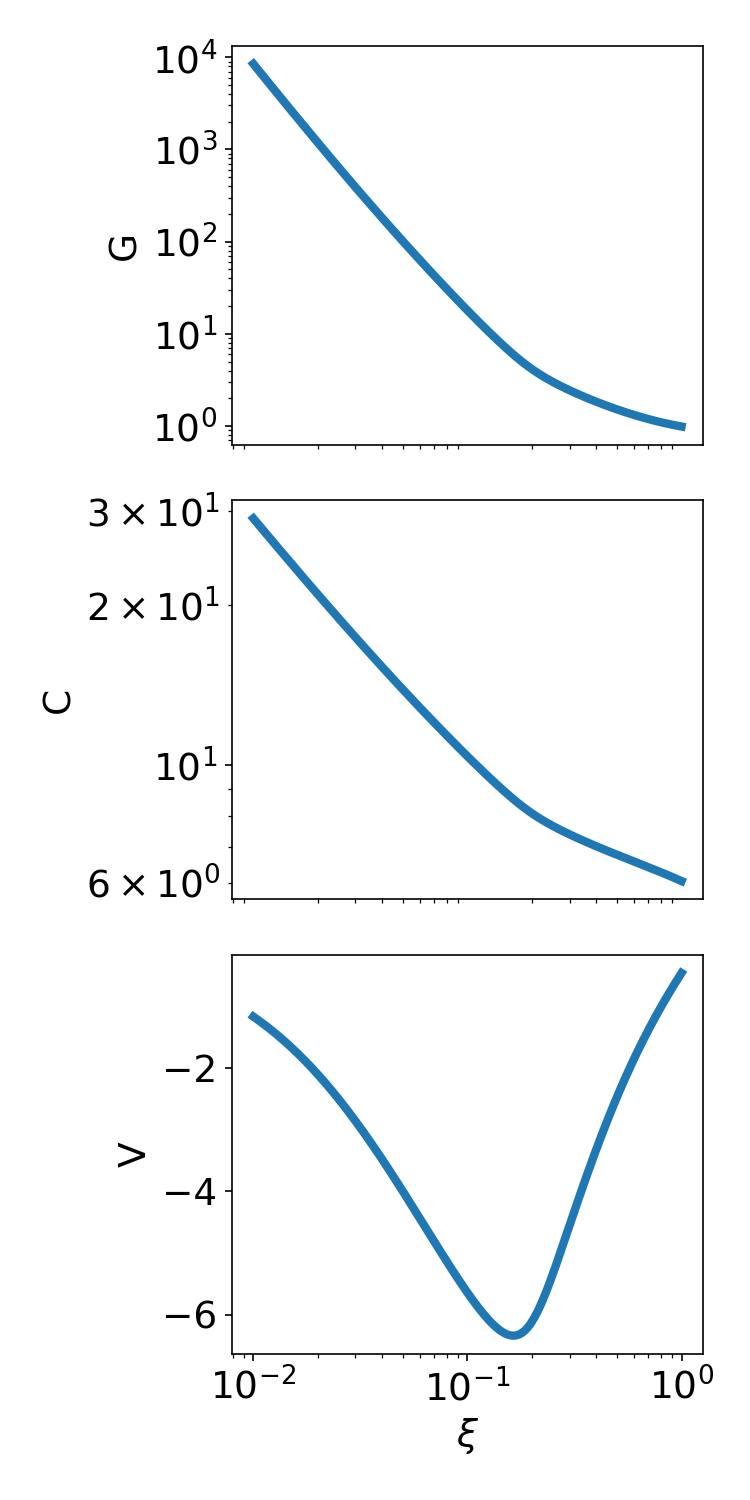}
\caption{Hydrodynamic profiles for $\gamma = 4/3$ and $\chi = 0.1$.} \label{fig:profile_1}
\end{figure}

\begin{figure}
    \centering
    \includegraphics[width=\linewidth]{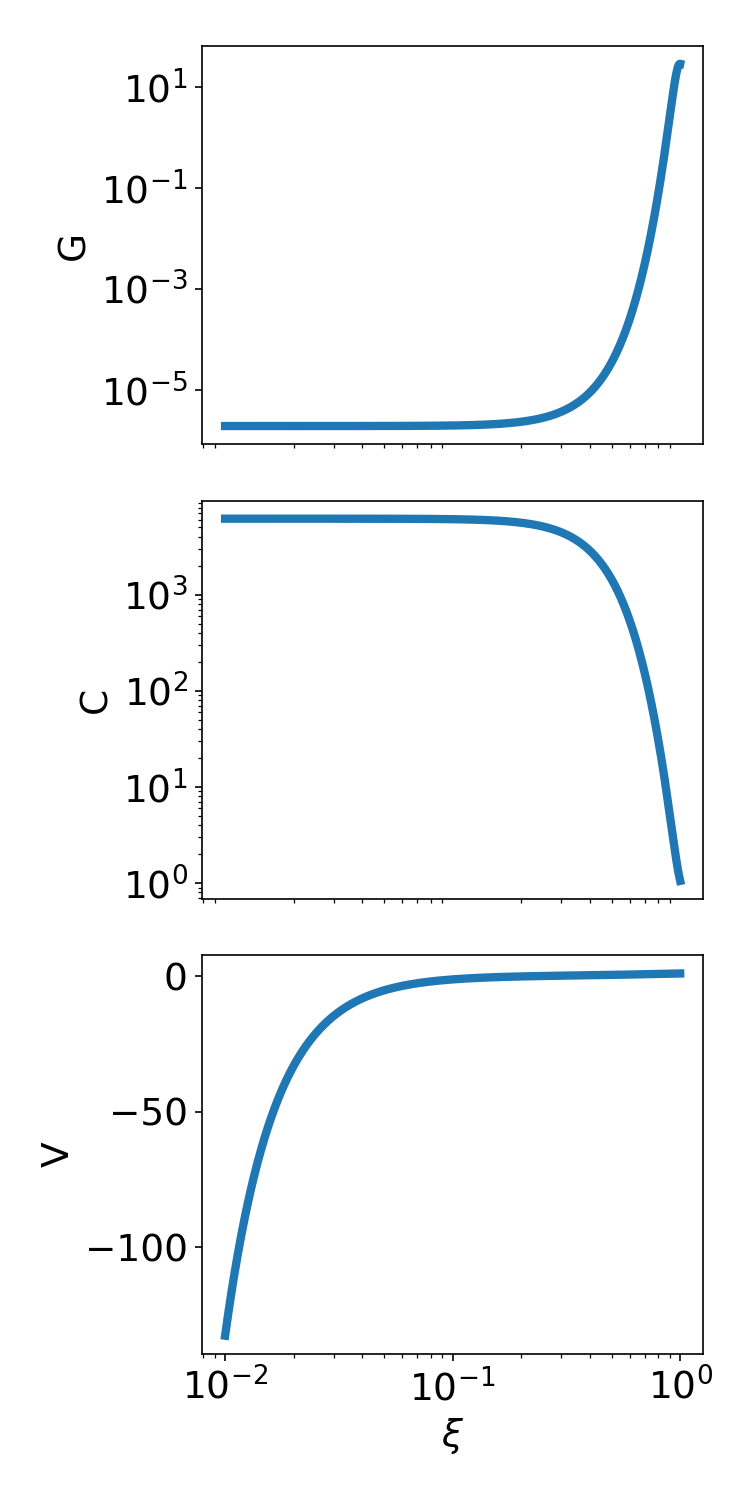}
    \caption{Hydrodynamic profiles for $\gamma = 1.01$ and $\chi = 0.1$.}
    \label{fig:profile_2}
\end{figure}

Using the dimensionless differential equations and the boundary conditions, one can numerically integrate to obtain the dimensionless hydrodynamic profiles. An example can be seen in figure \ref{fig:profile_1}. This example shows that the density peaks at small radii. As the adiabatic index is lowered, the peak shift toward the shock front, as can be seen in figure \ref{fig:profile_2}.

With the hydrodynamic profiles, we can use equation \ref{eq:energy_equation} to calculate the energy contained in the explosion. We can express this energy in dimensionless form using the virial parameter $\epsilon = E/4 \pi k G M$. In these units, the contribution of the potential energy of the ambient fluid prior to the explosion is unity (up to a logarithmic factor). 

\begin{figure}
    \centering
    \includegraphics[width=\linewidth]{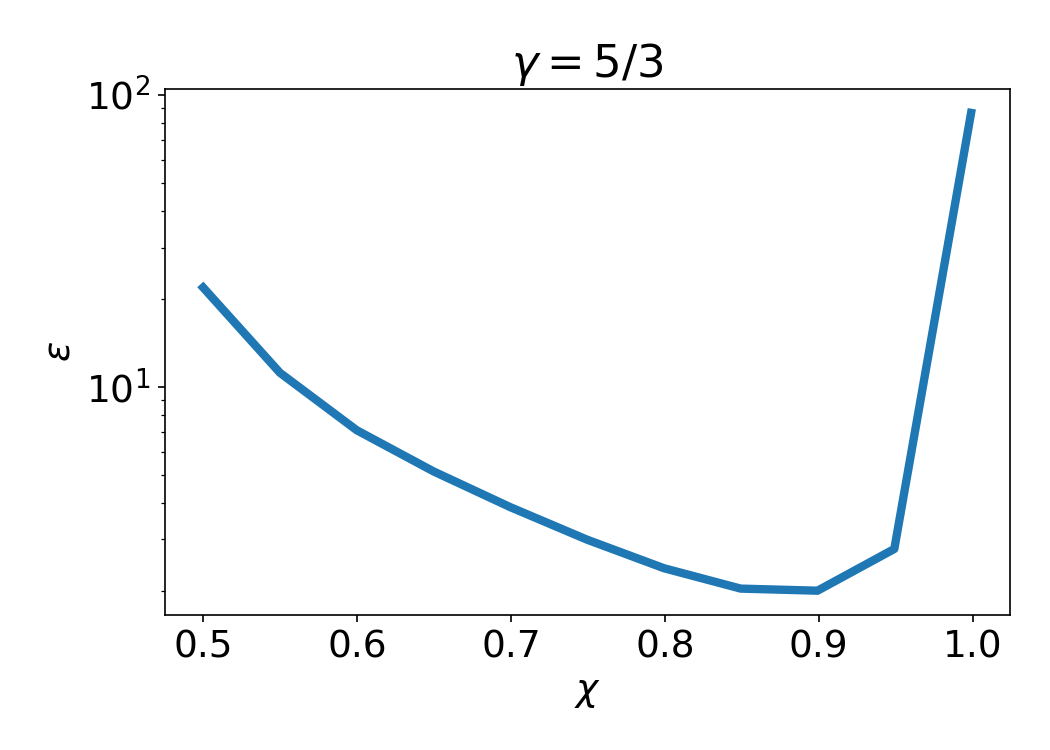}
    \caption{A plot of the virial parameter $\epsilon$ as a function of the radius ratio $\chi$ for a fixed adiabatic index. Minimum occurs for $\chi \approx 0.9$ and $\epsilon \approx 2$. The shock collapses when the energy dips below this critical limit, and similarly the only physical solution are those where $\chi$ is above the critical value.}
    \label{fig:epsilon_vs_chi}
\end{figure}

Figure \ref{fig:epsilon_vs_chi} shows the virial parameter $\epsilon$ as a function of the radius ratio $\chi$ for a fixed adiabatic index $\gamma=5/3$. We find that the virial parameter $\epsilon$ attains a minimum for a finite value of $\chi$. We can intuitively understand why the figure would exhibit such behaviour. In the limit $\chi \rightarrow 1$ we reproduce the known result without gravity, where $\epsilon$ increases with $\chi$. In the opposite limit $\chi \rightarrow 0$, as the radius ratio $\chi$ decreases, the pressure inside the explosion has to increase to balance gravity, and so the energy increases as well in this limit. Hence it is not surprising that in between these two asymptotic limits there has to be at least one minimum. The consequence of this realisation is that below some critical value of the virial parameter no self similar solution exists. Instead, the shock wave stalls and collapses. Another related consequence is that no solutions exist below some critical value of $\chi$. 

\subsection{Comparison to Simulations}

\begin{figure}
    \centering
    \includegraphics[width=\linewidth]{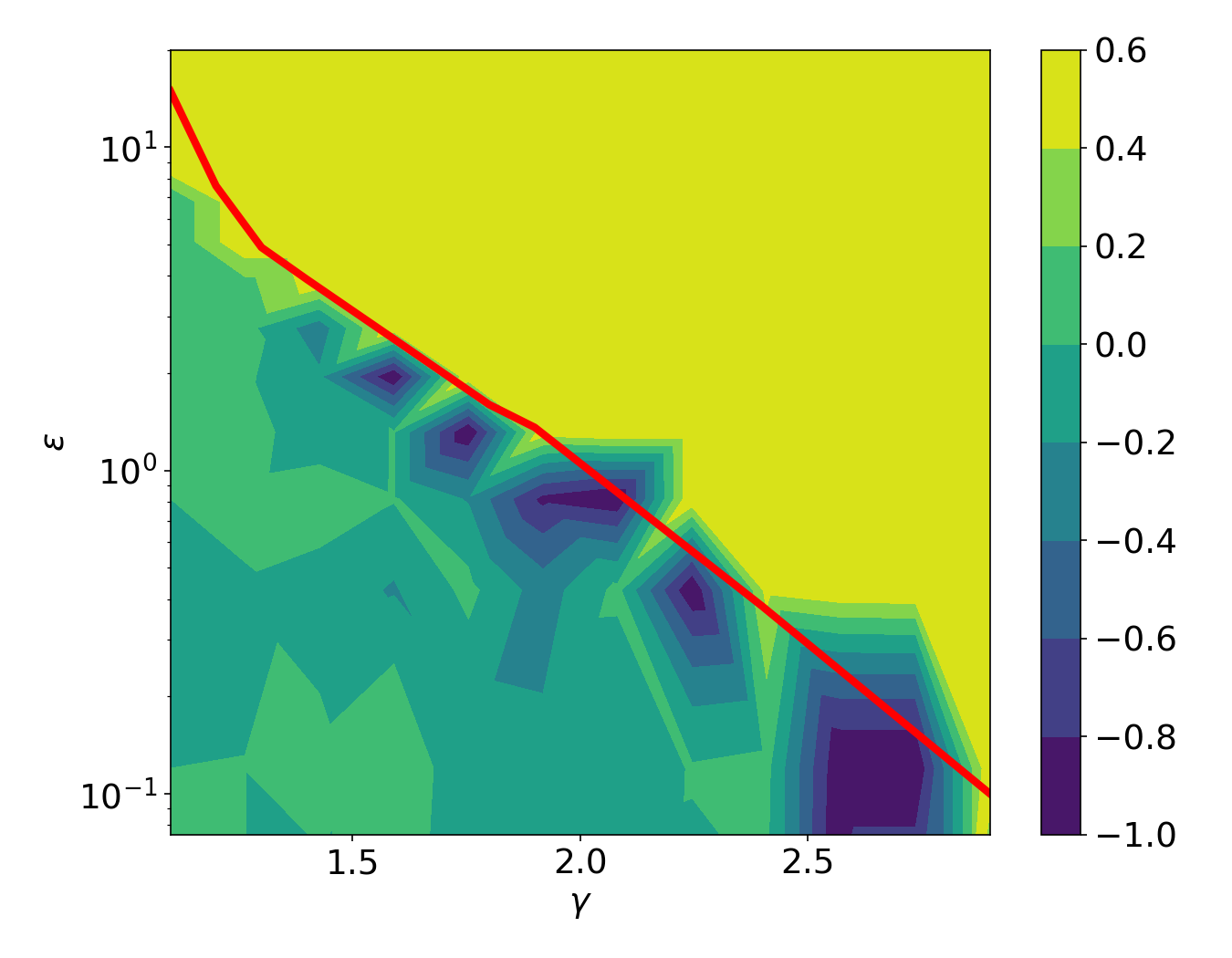}
    \caption{Results for the numerical simulations. For each value of the adiabatic index $\gamma$ and the virial parameter $\epsilon$ we ran a simulation, tracked the shock position at each time, fit to a power law $R = A t^{\alpha}$ and plotted $\alpha$. For enduring shock waves $\alpha = 2/3$, and if the shock stalls and collapses $\alpha$ quickly drops below this value. The red line is the theoretical threshold, obtained from the minimum of the $\epsilon \left(\chi\right)$ curve (see figure \ref{fig:epsilon_vs_chi}). The numerical results are in agreement with the theoretical predictions.}
    \label{fig:alpha_epsilon_gamma_map}
\end{figure}

To verify the ideas presented in the previous section, we ran a suite of one dimensional, spherically symmetric numerical simulations using the code RICH \citep{Yalinewich2015Rich:Mesh}. We set up the calculation with a logarithmic grid with 1000 cells between $r = 10^{-2}$ and $r = 10^3$. The initial density scales as $r^{-2}$, and the initial density is tiny throughout ($10^{-9}$) except for a small hot - spot in $r<10^{-1}$. The initial velocity was set to zero, and a gravitating mass was placed at $r=0$. All simulations ran till time $10^2$. We ran multiple simulations with different values of the adiabatic index $\gamma$ and different energy $\epsilon$. In each simulation we tracked the position of the shock wave and fit it to a trajectory of the form $R = A t^{\alpha}$. For an enduring shock we obtain $\alpha = 2/3$, but if the shock stalls and collapses the value of $\alpha$ quickly drops to lower values. In figure \ref{fig:alpha_epsilon_gamma_map} we plotted the numerical value of $\alpha$ for different values of $\gamma$ and $\epsilon$. In the same plot we also included a line that shows the theoretical minimal $\epsilon$ for every value of $\gamma$. The theoretical curve is in agreement with the numerical results, and it delineates the region of parameter space where enduring shocks form.

\section{Discussion} \label{sec:discussion}

In this work we developed a self similar solution for a spherical symmetric shock wave that propagates into a medium that is gravitationally collapsing into a point mass. We find that there is a critical amount of energy below which no such solutions exist, and the shock stalls and collapses. Moreover, we find that this energy threshold decreases as the adiabatic index increases. 

One can intuitively understand why increasing the adiabatic index would facilitate the propagation of the shock. The adiabatic index controls how material is distributed within the shocked region. In the limit $\gamma \rightarrow 1$ all the material swept up by the shock tends to accumulate close to the shock front, and as the adiabatic index increases, the material is spread out more evenly inside the shocked region. In the classical Sedov Taylor solution the velocity profile is almost homologous \citep{Landau1987FluidMechanics, Barenblatt2014ScalingAsymptotics}, i.e. linear in the radius, so the kinetic energy required to drive the shock increases as the adiabatic index decreases. Likewise, the potential gravitational energy also increases when the adiabatic index decreases. Therefore, as the adiabatic index increases, the material inside the shocked region is distributed in a more energy efficient way, and so the energy threshold is lower.

The solution presented here neglects many of the physical phenomena that take place during a supernova explosion, such as radiation, neutrino transport and turbulence. Nevertheless, it does reproduce the critical behaviour of supernova shock waves observed in previous studies \citep{Burrows1993AExplosions, Murphy2008CriteriaMechanism}. The novelty of this study is that it relates the structure of the shock to the explodability, from first principles.

Previous studies have shown that it is difficult to get spherically symmetric, one dimensional simulations to explode, but that it is easier in higher dimensions \citep{Bruenn2013AxisymmetricStars}. The main difference between one and multi dimensional simulation is turbulence. To first order approximation, turbulence doesn't add new energy into the system, so it is not immediately clear why it should improve the explodability. This study gives us a suggestion of what the solution might be: turbulence distributes the material in a more energy efficient way, like increasing the adiabatic index. To explore this idea further, we plan to develop a self similar solution that also includes a turbulent diffusion term. Another effect that can be included in a self similar solution is the constant neutrino heating from the proto neutron star, instead of the instantaneous release of energy assumed in this work. Finally, it is also worth considering higher order effects. One such effect is that a more uniform distribution of material inside the shocked region increases the optical depth, which in turn increases the amount of energy from absorbed neutrinos, and thus strengthens the shock.

\section*{Acknowledgements}

AY would like to thank Jeremiah Murphy, Chris Matzner, Mackenzie Warren and Sarah Gossan for the useful discussions. AY is supported by the Vincent  and  Beatrice  Tremaine  Fellowship. This work made use of the sympy \citep{Meurer2017SymPy:Python}, numpy \citep{Oliphant2006ANumPy} and matplotlib \citep{Hunter2007Matplotlib:Environment} python packages.

\section*{Data Availability}
The source code for the numerical simulation and documentation can be found on github at \url{https://github.com/bolverk/huji-rich}. The source code for the runs described in this work, as well as the output can be found on figshare \citep{Yalinewich2020GravitySimulation}.




\bibliographystyle{mnras}
\bibliography{main} 

\begin{thebibliography}{}
\makeatletter
\relax
\def\mn@urlcharsother{\let\do\@makeother \do\$\do\&\do\#\do\^\do\_\do\%\do\~}
\def\mn@doi{\begingroup\mn@urlcharsother \@ifnextchar [ {\mn@doi@}
  {\mn@doi@[]}}
\def\mn@doi@[#1]#2{\def\@tempa{#1}\ifx\@tempa\@empty \href
  {http://dx.doi.org/#2} {doi:#2}\else \href {http://dx.doi.org/#2} {#1}\fi
  \endgroup}
\def\mn@eprint#1#2{\mn@eprint@#1:#2::\@nil}
\def\mn@eprint@arXiv#1{\href {http://arxiv.org/abs/#1} {{\tt arXiv:#1}}}
\def\mn@eprint@dblp#1{\href {http://dblp.uni-trier.de/rec/bibtex/#1.xml}
  {dblp:#1}}
\def\mn@eprint@#1:#2:#3:#4\@nil{\def\@tempa {#1}\def\@tempb {#2}\def\@tempc
  {#3}\ifx \@tempc \@empty \let \@tempc \@tempb \let \@tempb \@tempa \fi \ifx
  \@tempb \@empty \def\@tempb {arXiv}\fi \@ifundefined
  {mn@eprint@\@tempb}{\@tempb:\@tempc}{\expandafter \expandafter \csname
  mn@eprint@\@tempb\endcsname \expandafter{\@tempc}}}

\bibitem[\protect\citeauthoryear{Barenblatt}{Barenblatt}{2014}]{Barenblatt2014ScalingAsymptotics}
Barenblatt G.~I.,  2014, {Scaling, self-similarity, and intermediate
  asymptotics}.
Cambridge University Press, \mn@doi{10.1017/CBO9781107050242}, \url
  {/core/books/scaling-selfsimilarity-and-intermediate-asymptotics/3B56096C3B7E822794C81B51F7370B82}

\bibitem[\protect\citeauthoryear{Brandt, Burrows, Ott  \& Livne}{Brandt
  et~al.}{2010}]{Brandt2010ResultsTransport}
Brandt T.~D.,  Burrows A.,  Ott C.~D.,   Livne E.,  2010, \mn@doi
  [Astrophysical Journal] {10.1088/0004-637X/728/1/8}, 728, 8

\bibitem[\protect\citeauthoryear{Bruenn et~al.,}{Bruenn
  et~al.}{2013}]{Bruenn2013AxisymmetricStars}
Bruenn S.~W.,  et~al., 2013, \mn@doi [Astrophysical Journal Letters]
  {10.1088/2041-8205/767/1/L6}, 767, 6

\bibitem[\protect\citeauthoryear{Buras, Rampp, Janka  \& Kifonidis}{Buras
  et~al.}{2005}]{Buras2005Two-dimensionalStar}
Buras R.,  Rampp M.,  Janka H.~T.,   Kifonidis K.,  2005, \mn@doi [Astronomy
  and Astrophysics] {10.1051/0004-6361:20053783}, 447, 1049

\bibitem[\protect\citeauthoryear{Burrows}{Burrows}{2012}]{Burrows2012PerspectivesTheory}
Burrows A.,  2012, \mn@doi [Reviews of Modern Physics]
  {10.1103/RevModPhys.85.245}, 85, 245

\bibitem[\protect\citeauthoryear{Burrows \& Goshy}{Burrows \&
  Goshy}{1993}]{Burrows1993AExplosions}
Burrows A.,  Goshy J.,  1993, \mn@doi [The Astrophysical Journal]
  {10.1086/187074}, 416, L75

\bibitem[\protect\citeauthoryear{Burrows, Radice  \& Vartanyan}{Burrows
  et~al.}{2019}]{Burrows2019Three-dimensionalStars}
Burrows A.,  Radice D.,   Vartanyan D.,  2019, \mn@doi [Monthly Notices of the
  Royal Astronomical Society] {10.1093/mnras/stz543}, 485, 3169

\bibitem[\protect\citeauthoryear{Cabez{\'{o}}n, Pan, Liebend{\"{o}}rfer,
  Kuroda, Ebinger, Heinimann, Perego  \& Thielemann}{Cabez{\'{o}}n
  et~al.}{2018}]{Cabezon2018Core-collapseProject}
Cabez{\'{o}}n R.~M.,  Pan K.~C.,  Liebend{\"{o}}rfer M.,  Kuroda T.,  Ebinger
  K.,  Heinimann O.,  Perego A.,   Thielemann F.~K.,  2018, \mn@doi [Astronomy
  and Astrophysics] {10.1051/0004-6361/201833705}, 619, A118

\bibitem[\protect\citeauthoryear{Chevalier}{Chevalier}{1989}]{Chevalier1989NeutronSupernova}
Chevalier R.~A.,  1989, \mn@doi [The Astrophysical Journal] {10.1086/168066},
  346, 847

\bibitem[\protect\citeauthoryear{Couch \& Ott}{Couch \&
  Ott}{2014}]{Couch2014TheExplosions}
Couch S.~M.,  Ott C.~D.,  2014, \mn@doi [Astrophysical Journal]
  {10.1088/0004-637X/799/1/5}, 799, 5

\bibitem[\protect\citeauthoryear{Couch, Warren  \& O’Connor}{Couch
  et~al.}{2020}]{Couch2020SimulatingDimension}
Couch S.~M.,  Warren M.~L.,   O’Connor E.~P.,  2020, \mn@doi [The
  Astrophysical Journal] {10.3847/1538-4357/ab609e}, 890, 127

\bibitem[\protect\citeauthoryear{Coughlin, Quataert  \& Ro}{Coughlin
  et~al.}{2018}]{Coughlin2018WeakSupernovae}
Coughlin E.~R.,  Quataert E.,   Ro S.,  2018, \mn@doi [The Astrophysical
  Journal] {10.3847/1538-4357/aad198}, 863, 158

\bibitem[\protect\citeauthoryear{Dolence, Burrows  \& Zhang}{Dolence
  et~al.}{2015}]{Dolence2015Two-DimensionalTransport}
Dolence J.~C.,  Burrows A.,   Zhang W.,  2015, \mn@doi [Astrophysical Journal
  Letters] {10.1088/0004-637X/800/1/10}, 800, 10

\bibitem[\protect\citeauthoryear{Fryer \& Warren}{Fryer \&
  Warren}{2002}]{Fryer2002ModelingDimensions}
Fryer C.~L.,  Warren M.~S.,  2002, \mn@doi [The Astrophysical Journal]
  {10.1086/342258}, 574, L65

\bibitem[\protect\citeauthoryear{Hunter}{Hunter}{2007}]{Hunter2007Matplotlib:Environment}
Hunter J.~D.,  2007, \mn@doi [Computing in Science and Engineering]
  {10.1109/MCSE.2007.55}, 9, 90

\bibitem[\protect\citeauthoryear{Janka, Langanke, Marek, Mart{\'{i}}nez-Pinedo
  \& M{\"{u}}ller}{Janka et~al.}{2007}]{Janka2007TheorySupernovae}
Janka H.~T.,  Langanke K.,  Marek A.,  Mart{\'{i}}nez-Pinedo G.,   M{\"{u}}ller
  B.,  2007, {Theory of core-collapse supernovae},
  \mn@doi{10.1016/j.physrep.2007.02.002}, \url
  {https://ui.adsabs.harvard.edu/abs/2007PhR...442...38J/abstract}

\bibitem[\protect\citeauthoryear{Kazhdan \& Murzina}{Kazhdan \&
  Murzina}{1994}]{Kazhdan1994Self-similarMass}
Kazhdan Y.~M.,  Murzina M.,  1994, \mn@doi [Monthly Notices of the Royal
  Astronomical Society] {10.1093/mnras/270.2.351}, 270, 351

\bibitem[\protect\citeauthoryear{Landau \& Lifshitz}{Landau \&
  Lifshitz}{1987}]{Landau1987FluidMechanics}
Landau L.~D.,  Lifshitz E.~M.,  1987, {Fluid Mechanics},
  \mn@doi{10.1007/b138775}

\bibitem[\protect\citeauthoryear{Lemos \& Lynden-Bell}{Lemos \&
  Lynden-Bell}{1989}]{Lemos1989AGravity}
Lemos J. P.~S.,  Lynden-Bell D.,  1989, \mn@doi [Monthly Notices of the Royal
  Astronomical Society] {10.1093/mnras/240.2.317}, 240, 317

\bibitem[\protect\citeauthoryear{Livne, Burrows, Walder, Lichtenstadt  \&
  Thompson}{Livne et~al.}{2003}]{Livne2003Two-dimensionalContext}
Livne E.,  Burrows A.,  Walder R.,  Lichtenstadt I.,   Thompson T.~A.,  2003,
  \mn@doi [The Astrophysical Journal] {10.1086/421012}, 609, 277

\bibitem[\protect\citeauthoryear{Mabanta \& Murphy}{Mabanta \&
  Murphy}{2018}]{Mabanta2018HowExplosions}
Mabanta Q.~A.,  Murphy J.~W.,  2018, \mn@doi [The Astrophysical Journal]
  {10.3847/1538-4357/aaaec7}, 856, 22

\bibitem[\protect\citeauthoryear{Mabanta, Murphy  \& Dolence}{Mabanta
  et~al.}{2019}]{Mabanta2019Convection-aidedValidation}
Mabanta Q.~A.,  Murphy J.~W.,   Dolence J.~C.,  2019, \mn@doi [The
  Astrophysical Journal] {10.3847/1538-4357/ab4bcc}, 887, 43

\bibitem[\protect\citeauthoryear{Melson, Janka  \& Marek}{Melson
  et~al.}{2015}]{Melson2015Neutrino-drivenConvection}
Melson T.,  Janka H.~T.,   Marek A.,  2015, \mn@doi [Astrophysical Journal
  Letters] {10.1088/2041-8205/801/2/L24}, 801, L24

\bibitem[\protect\citeauthoryear{Meurer et~al.,}{Meurer
  et~al.}{2017}]{Meurer2017SymPy:Python}
Meurer A.,  et~al., 2017, \mn@doi [PeerJ Computer Science]
  {10.7717/peerj-cs.103}, 3, e103

\bibitem[\protect\citeauthoryear{Mezzacappa, Liebend{\"{o}}rfer, Messer, Hix,
  Thielemann  \& Bruenn}{Mezzacappa
  et~al.}{2001}]{Mezzacappa2001SimulationMechanism}
Mezzacappa A.,  Liebend{\"{o}}rfer M.,  Messer O.~E.,  Hix W.~R.,  Thielemann
  F.~K.,   Bruenn S.~W.,  2001, \mn@doi [Physical Review Letters]
  {10.1103/PhysRevLett.86.1935}, 86, 1935

\bibitem[\protect\citeauthoryear{M{\"{u}}ller}{M{\"{u}}ller}{2017}]{Muller2017TheMechanism}
M{\"{u}}ller B.,  2017, \mn@doi [Proceedings of the International Astronomical
  Union] {10.1017/S1743921317002575}, 12, 17

\bibitem[\protect\citeauthoryear{Murphy \& Burrows}{Murphy \&
  Burrows}{2008}]{Murphy2008CriteriaMechanism}
Murphy J.~W.,  Burrows A.,  2008, \mn@doi [The Astrophysical Journal]
  {10.1086/592214}, 688, 1159

\bibitem[\protect\citeauthoryear{Murphy \& Dolence}{Murphy \&
  Dolence}{2015}]{Murphy2015AnExplosions}
Murphy J.~W.,  Dolence J.~C.,  2015, \mn@doi [The Astrophysical Journal]
  {10.3847/1538-4357/834/2/183}, 834, 183

\bibitem[\protect\citeauthoryear{Murphy \& Meakin}{Murphy \&
  Meakin}{2011}]{Murphy2011ASupernovae}
Murphy J.~W.,  Meakin C.,  2011, \mn@doi [Astrophysical Journal]
  {10.1088/0004-637X/742/2/74}, 742, 74

\bibitem[\protect\citeauthoryear{O'Connor, Horowitz, Lin  \& Couch}{O'Connor
  et~al.}{2017}]{OConnor2017Core-CollapseEOS}
O'Connor E.,  Horowitz C.~J.,  Lin Z.,   Couch S.,  2017, \mn@doi [Proceedings
  of the International Astronomical Union] {10.1017/S1743921317004586}, 12, 107

\bibitem[\protect\citeauthoryear{O'Connor et~al.,}{O'Connor
  et~al.}{2018}]{OConnor2018GlobalSymmetry}
O'Connor E.,  et~al., 2018, \mn@doi [Journal of Physics G: Nuclear and Particle
  Physics] {10.1088/1361-6471/aadeae}, 45, 104001

\bibitem[\protect\citeauthoryear{Oliphant}{Oliphant}{2006}]{Oliphant2006ANumPy}
Oliphant T.~E.,  2006, {A guide to NumPy}

\bibitem[\protect\citeauthoryear{Ott, Burrows, Dessart  \& Livne}{Ott
  et~al.}{2008}]{Ott20082DCores}
Ott C.~D.,  Burrows A.,  Dessart L.,   Livne E.,  2008, \mn@doi [The
  Astrophysical Journal] {10.1086/591440}, 685, 1069

\bibitem[\protect\citeauthoryear{Ott et~al.,}{Ott
  et~al.}{2013}]{Ott2013General-relativisticSupernovae}
Ott C.~D.,  et~al., 2013, \mn@doi [Astrophysical Journal]
  {10.1088/0004-637X/768/2/115}, 768, 115

\bibitem[\protect\citeauthoryear{Pan, Liebend{\"{o}}rfer, Hempel  \&
  Thielemann}{Pan et~al.}{2016}]{Pan2016TWO-DIMENSIONALTRANSPORT}
Pan K.-C.,  Liebend{\"{o}}rfer M.,  Hempel M.,   Thielemann F.-K.,  2016,
  \mn@doi [The Astrophysical Journal] {10.3847/0004-637x/817/1/72}, 817, 72

\bibitem[\protect\citeauthoryear{Radice, Abdikamalov, Ott, M{\"{o}}sta, Couch
  \& Roberts}{Radice et~al.}{2018}]{Radice2018TurbulenceSupernovae}
Radice D.,  Abdikamalov E.,  Ott C.~D.,  M{\"{o}}sta P.,  Couch S.~M.,
  Roberts L.~F.,  2018, {Turbulence in core-collapse supernovae},
  \mn@doi{10.1088/1361-6471/aab872}, \url
  {https://ui.adsabs.harvard.edu/abs/2018JPhG...45e3003R/abstract}

\bibitem[\protect\citeauthoryear{Rogers}{Rogers}{1957}]{Rogers1957AnalyticDensity.}
Rogers M.~H.,  1957, \mn@doi [The Astrophysical Journal] {10.1086/146323}, 125,
  478

\bibitem[\protect\citeauthoryear{Sakashita}{Sakashita}{1974}]{Sakashita1974SimilarityFlow}
Sakashita S.,  1974, \mn@doi [Astrophysics and Space Science]
  {10.1007/BF00642634}, 26, 183

\bibitem[\protect\citeauthoryear{Takiwaki, Kotake  \& Suwa}{Takiwaki
  et~al.}{2011}]{Takiwaki2011Three-dimensionalTransport}
Takiwaki T.,  Kotake K.,   Suwa Y.,  2011, \mn@doi [Astrophysical Journal]
  {10.1088/0004-637X/749/2/98}, 749, 98

\bibitem[\protect\citeauthoryear{Takiwaki, Kotake  \& Suwa}{Takiwaki
  et~al.}{2014}]{Takiwaki2014ASupernovae}
Takiwaki T.,  Kotake K.,   Suwa Y.,  2014, \mn@doi [Astrophysical Journal]
  {10.1088/0004-637X/786/2/83}, 786, 83

\bibitem[\protect\citeauthoryear{Ugliano, Janka, Marek  \& Arcones}{Ugliano
  et~al.}{2012}]{Ugliano2012Progenitor-explosionProgenitors}
Ugliano M.,  Janka H.~T.,  Marek A.,   Arcones A.,  2012, \mn@doi
  [Astrophysical Journal] {10.1088/0004-637X/757/1/69}, 757, 69

\bibitem[\protect\citeauthoryear{Van~Dyk}{Van~Dyk}{2017}]{VanDyk2017TheProgenitors}
Van~Dyk S.~D.,  2017, \mn@doi [Philosophical Transactions of the Royal Society
  A: Mathematical, Physical and Engineering Sciences] {10.1098/rsta.2016.0277},
  375, 20160277

\bibitem[\protect\citeauthoryear{Woosley \& Heger}{Woosley \&
  Heger}{2007}]{Woosley2007NucleosynthesisMetallicity}
Woosley S.~E.,  Heger A.,  2007, {Nucleosynthesis and remnants in massive stars
  of solar metallicity}, \mn@doi{10.1016/j.physrep.2007.02.009}, \url
  {https://ui.adsabs.harvard.edu/abs/2007PhR...442..269W/abstract}

\bibitem[\protect\citeauthoryear{Yalinewich}{Yalinewich}{2020}]{Yalinewich2020GravitySimulation}
Yalinewich A.,  2020, {gravity explosion simulation},
  \mn@doi{10.6084/m9.figshare.12860774}, \url
  {https://figshare.com/articles/dataset/gravity_explosion_simulation/12860774/1}

\bibitem[\protect\citeauthoryear{Yalinewich, Steinberg  \& Sari}{Yalinewich
  et~al.}{2015}]{Yalinewich2015Rich:Mesh}
Yalinewich A.,  Steinberg E.,   Sari R.,  2015, \mn@doi [Astrophysical Journal,
  Supplement Series] {10.1088/0067-0049/216/2/35}, 216

\bibitem[\protect\citeauthoryear{Zel'dovich \& Raizer}{Zel'dovich \&
  Raizer}{1967}]{Zeldovich1967PhysicsPhenomena}
Zel'dovich Y.~B.,  Raizer Y.~P.,  1967, {Physics of shock waves and
  high-temperature hydrodynamic phenomena}.
Dover

\makeatother
\end{thebibliography}








\bsp	
\label{lastpage}
\end{document}